\documentclass[aps,twocolumn,showpacs,preprintnumbers,prb,superscriptaddress]{revtex4-1}
\usepackage{graphicx}
\usepackage{amsmath}
\usepackage{amssymb}

\usepackage{color}

\newcommand{\R}{\mathbf{r}}

\newcommand{\be}{\begin{equation}}
\newcommand{\ee}{\end{equation}}
\newcommand{\bea}{\begin{eqnarray}}
\newcommand{\eea}{\end{eqnarray}}
\newcommand{\bean}{\begin{eqnarray*}}
\newcommand{\eean}{\end{eqnarray*}}

\begin{document}

\title{Jellium-with-gap model applied to semilocal kinetic functionals}
\author{Lucian A. Constantin}
\affiliation{Center for Biomolecular Nanotechnologies @UNILE, Istituto Italiano di Tecnologia, Via Barsanti, 
I-73010 Arnesano, Italy}
\author{Eduardo Fabiano}
\affiliation{Institute for Microelectronics and Microsystems (CNR-IMM), Via Monteroni, Campus Unisalento, 73100 Lecce, Italy}
\affiliation{Center for Biomolecular Nanotechnologies @UNILE, Istituto Italiano di Tecnologia, Via Barsanti, 
I-73010 Arnesano, Italy}
\author{Szymon \'Smiga}
\affiliation{Istituto Nanoscienze-CNR, Italy}
\affiliation{Institute of Physics, Faculty of Physics, Astronomy and Informatics, Nicolaus Copernicus
University, Grudziadzka 5, 87-100 Torun, Poland}
\affiliation{Center for Biomolecular Nanotechnologies @UNILE, Istituto Italiano di Tecnologia, Via Barsanti,
I-73010 Arnesano, Italy}
\author{Fabio Della Sala}
\affiliation{Institute for Microelectronics and Microsystems (CNR-IMM), Via Monteroni, Campus Unisalento, 73100 Lecce, Italy}
\affiliation{Center for Biomolecular Nanotechnologies @UNILE, Istituto Italiano di Tecnologia, Via Barsanti, 
I-73010 Arnesano, Italy}
\date{\today}

\begin{abstract}
We investigate a highly-nonlocal generalization of the Lindhard function, given by the jellium-with-gap model.
We find a band-gap-dependent gradient expansion of the kinetic energy, which
performs noticeably well for large atoms. 
Using the static linear response theory and the simplest semilocal model for the local band gap, we derive a
non-empirical generalized gradient approximation (GGA) of the kinetic energy.
This GGA kinetic energy functional is remarkably accurate for 
the description of
weakly interacting molecular systems within the subsystem formulation of Density Functional Theory.
\end{abstract}


\maketitle

\section{Introduction}
Density Functional Theory (DFT) \cite{yang_parr_book,dreiz_book} is the most used 
computational method for electronic structure calculations of  molecular and extended systems, 
providing the highest accuracy/computational cost ratio. 
In the conventional DFT formalism, the Kohn-Sham (KS) scheme \cite{kohn1965self}, the ground-state electronic density 
$n(\R)$ is determined from a set of auxiliary KS orbitals ($\phi_i(\R)$): the KS-DFT method is exact but for the 
approximations of the exchange-correlation (XC) functional.
However, for large scale calculations, the computational cost of KS-DFT
becomes unaffordable, as one needs to compute all the occupied KS 
orbitals in order to construct the density as $n(\R)=\sum_i^{occ.} f_i |\phi_i(\R)|^2$, where $f_i$ is the occupation number (2, 
for closed-shell systems). 

Among other linear scaling methods \cite{yang91,god99,bowler02,kuss13}, two DFT methods are attracting strong 
interest:
i) In the orbital-free version of DFT (OF-DFT) \cite{sma94,wang2002orbital,gavini07,wesobook,chen16},  $n(\R)$ can be 
computed directly via the 
Euler equation \cite{yang_parr_book}, without the  need of KS orbtials;
ii) In the subsystem version of DFT (Sub-DFT) \cite{cortona,wesorev,pavanello15,wesochemrev}, also known as 
Frozen-Density-Embedding (FDE), 
$n(\R)$ is computed as the sum of the electronic densities of several (smaller) subsystems	
in which the total system is partitioned, which can be computed simultaneously.
Both approaches allow in principle calculations of large systems, but the
final accuracy depends directly on the approximations of the non-interacting
kinetic energy (KE) functional $T_s$ (which 
are definitely more important than the ones for the XC energy, that are also present in standard KS 
calculations).
We recall that the exact KS KE functional is:
\begin{equation}
T_s^{exact}=\frac{1}{2}\sum_i^{occ.} \int f_i |\nabla \phi_i({\bf r})|^2 {\rm d^3}{\bf r} .
\end{equation}
%
Thus the KE is explicitly known only as a function of $\phi_i$ but not as a 
functional of $n$.
 
On the other hand, in Sub-DFT the interaction between the subsystems is taken into account via the so called embedding 
potentials \cite{wesorev,pavanello15,wesochemrev}, 
which depends on the non-additive-KE: in the case of just two subsystems (A and B) it is 
$T_s^{nadd}[n_A;n_B]=T_s[n_A+n_B]-T_s[n_A]-T_s[n_B]$.

The development of an accurate approximation of $T_s[n]$ (and/or
$T_s^{nadd}[n_A;n_B]$) is one of the biggest DFT  
challenges \cite{kara12,carter12,lastra08}.
Nowadays, the most sophisticated KE approximations have been constructed to be exact for the linear response of 
jellium model, by incorporating the Lindhard function in their fully non-local expressions 
\cite{wang2002orbital,wang1998orbital,ho2008analytic,garcia1996nonlocal,garcia1998nonlocal,garcia2008approach}.
We recall that the Lindhard function 
\cite{lindhard1954properties,wang2002orbital}
\begin{equation}
F^{Lind}=\left( \frac{1}{2}+\frac{1-\eta^2}{4\eta}\ln \left| \frac{1+\eta}{1-\eta} \right| \right)^{-1},
\label{eq1}
\end{equation}
where $\eta=k/(2k_F)$ is the dimensionless momentum ($k_F=(3\pi^2 n)^{1/3}$ being the Fermi 
wave vector of the jellium model with the constant density $n$), 
is related to the Jellium density response $\chi^{Jell}$ via\cite{wang2002orbital}
\begin{equation}
\label{chif}
-\frac{1}{\chi^{Jell.}}=\frac{\pi^2}{k_F} F^{Lind} \, .
\end{equation}

The non-local KE functionals based on the  Lindhard function are accurate for simple metals 
 where the nearly-free electron gas is an excellent model
 but they can not describe well
semiconductors and insulators, where the density response function behaves as
  \cite{pick1970microscopic,huang2010nonlocal}
\begin{equation}
-\frac{1}{\chi^{Semic.}(k)}  \underset{k\rightarrow 0}{\longrightarrow}    \frac{b}{k^2},
\label{eqllw1}
\end{equation}
with $b$ being positive and material-dependent. Several KE functionals have
been constructed
to improve the description of semiconductors \cite{huang2010nonlocal,shin2014enhanced}, but Eq. (\ref{eqllw1}) has not been 
explicitly used in their expressions due to the lack of a sophisticated analytical form that can recover both 
the Lindhard function and Eq. (\ref{eqllw1}).

In this article, we will investigate the generalization 
of the Lindhard function for the jellium-with-gap model which satisfies Eq. (\ref{eqllw1}).


The  jellium-with-gap model\cite{rey1998virtual}, was developed outside the KS framework,
using perturbation theory to take into account the band
gap energy. This model was used to have qualitative and quantitative insight for semiconductors 
\cite{callaway1959correlation,penn1962wave,srinivasan1969microscopic,levine1982new,tsolakidis2004effect},
to develop an XC kernel for the optical properties of materials 
\cite{trevisanutto2013optical}, and to construct accurate correlation energy functionals for the ground-state DFT 
\cite{rey1998virtual,krieger1999electron,krieger2001density,toulouse2002validation,
toulouse2002new,fabiano2014generalized}.
We will show that the Lindhard function for the jellium-with-gap model ($F^{GAP}$), previously 
introduced by Levine and Louie \cite{levine1982new} in a different
context (dielectric constant and XC potential), 
may be seen as a sophisticated analytical form suitable for KE approximations. 

The article is organized as follow:

In Section II we discuss the properties of $F^{GAP}$, we derive its (band-gap-dependent) KE gradient expansion, 
and we assess it for large atoms. By using a local gap model, we propose
a simple KE gradient expansion that it is very accurate for the semiclassical atom theory.  

In Section III we discuss the implications of this result in
DFT by constructing a simple KE functional at the
Generalized Gradient Approximation (GGA) level of theory based on the gradient expansion of
the jellium-with-gap model. 
GGA KE functionals are computationally very efficient and play a key 
role for the simulation of large systems. 
We mention that the development of semilocal KE functionals is nowadays an active field 
\cite{constantin2011semiclassical,laricchia2011generalized,kara13,laricchia2013laplacian,
borgo2013density,borgo14,kinairy14,alpha,cancioJCP16,smiga2017}.

Finally, in Section IV we summarize our results.

\section{Theory}

\subsection{Properties and gradient expansions for the jellium model}
For the conventional infinite jellium model, the Lindhard function behaves as:
\begin{eqnarray}
\label{eq2} 
F^{Lind}& \rightarrow & 1+ \frac{1}{3}\eta^2+\frac{8}{45}\eta^4+\mathcal{O}(\eta^6),\;\;\rm{for}\;\;
\eta\rightarrow 0, \\
%
\label{eq3}
F^{Lind} & \rightarrow & 3\eta^2 -\frac{3}{5}-\frac{24}{175}\frac{1}{\eta^2}+\mathcal{O}(\eta^{-4}),
\;\;\rm{for}\;\;
\eta\rightarrow \infty.
\end{eqnarray}
Equation (\ref{eq2}) contains important physics that has been used in the construction 
of semilocal
KE density functionals \cite{wang2002orbital}. 
Thus, 
the KE gradient expansion which recovers the first three terms in the right hand side of Eq. (\ref{eq2})
can be easily derived \cite{wang2002orbital} (see also Eqs. (\ref{gap4exp}) and
(\ref{eq14}) in section II-C and the corresponding discussion). It is  
\begin{equation}
T_s^{Lind4}[n]=\int d\R\, \tau^{TF}\left(1+\frac{5}{27}s^2+\frac{8}{81}q^2\right),
\label{eq4}
\end{equation}
where $\tau^{TF}=\frac{3}{10}(3\pi^{2})^{2/3}n^{5/3}$ is the Thomas-Fermi KE density 
\cite{thomas1927calculation,fermi1927metodo}, which is exact for the jellium model, and 
$s=|\nabla n|/[2 k_F n]$, $q=\nabla n^{2}/[4(3\pi^{2})^{2/3}n^{5/3}]$ are the reduced 
gradient and Laplacian, respectively.
Equation (\ref{eq4}) resembles the second-order gradient expansion \cite{Kirz57} (GE2)
$$T_s^{GE2}[n]=\int d\R \tau^{TF}(1+\frac{5}{27}s^2)$$
(derived also within the linear response of the jellium model), 
as well as the fourth-order gradient expansion 
\cite{hodges1973quantum,brack1976extended,laricchia2013laplacian}
of the KE
$$T_s^{GE4}[n]=\int d\R \tau^{TF}(1+\frac{5}{27}s^2+\frac{8}{81}q^2-\frac{1}{9}s^2 q +\frac{8}{243}s^4),$$ 
with the exception of the terms $\propto s^2 q$, $\propto s^4$, which are beyond the linear response.

Note that $F^{Lind}(\eta=0)=1$ is the leading term in the expansion of
  Eq. (5) and 
it corresponds to the 
Thomas-Fermi local density approximation, whose linear response in the wave vector space is just the Fourier 
transform of the second-functional derivative, i.e. $\delta^2 T_s^{TF}/\delta n(\R)\delta n(\R')\sim 
k_F^{-1}\delta(\R-\R')$.  
We recall that the limit $\eta=0$ is very powerful, being also used in the construction of the adiabatic local 
density approximation (ALDA) XC kernel of 
the linear response time-dependent DFT \cite{burke2005time,constantin2007simple}.

\subsection{Properties of the Lindhard function for the jellium-with-gap model}
Levine and Louie \cite{levine1982new} proposed
the density-response function $\chi^{GAP}(k,\omega)$ of the jellium-with-gap model, and the corresponding [i.e. from
Eq. (\ref{chif})]
Lindhard function for jellium-with-gap model is
\begin{eqnarray}
&& 1/F^{GAP}=\frac{1}{2}-\frac{\Delta(\arctan(\frac{4\eta+4\eta^2}{\Delta})+
\arctan(\frac{4\eta-4\eta^2}{\Delta}))}{8\eta}\nonumber+ \\
&& \;\;\; + (\frac{\Delta^2}{128\eta^3}+\frac{1}{8\eta}-\frac{\eta}{8})\ln(\frac{\Delta^2+(4\eta+4\eta^2)^2}
{\Delta^2+(4\eta-4\eta^2)^2}),
\label{eq6}
\end{eqnarray}
where $\Delta=2E_g/k_F^2$ and $E_g$ is the gap.

For a given $\Delta$, a series expansion of $F^{GAP}$ for $\eta\rightarrow 0$ gives: 
\begin{eqnarray}
&& F^{GAP}\longrightarrow \frac{3\Delta^2}{16\eta^2}+\frac{9}{5}+
\frac{3}{175}\frac{175\Delta^2-192}{\Delta^2}\eta^2+ \nonumber \\
&&\;\;\; -\frac{64}{875}\frac{525\Delta^2-368}{\Delta^4}\eta^4+
\mathcal{O}(\eta^6)\;\;\rm{when}\;\;\eta\rightarrow 0 \; .
\label{eqe0}   
\end{eqnarray}
Thus, for any system with $\Delta >0$ we have that $F^{GAP}\propto \Delta^2 \eta^{-2}$.
This term is correct (see Eq. (4)) and it has been also used in the jellium-with-gap XC kernel 
\cite{trevisanutto2013optical}, 
which gives accurate optical absorption spectra of semiconductors and insulators. 
On the other hand, if we first perform a series expansion for  $\Delta\rightarrow 0$, and then
a series expansion for  $\eta\rightarrow 0$ we obtain:
\begin{eqnarray}
&& F^{GAP}\longrightarrow \left[1+\frac{1}{3}\eta^2+\frac{8}{45}\eta^4+...\right]+ \nonumber\\
&& \Delta \left[\frac{\pi}{8}\frac{1}{\eta}+\frac{\pi}{12}\eta+\frac{7\pi}{120}\eta^3+...\right]+ 
\Delta^2 \Bigg[\frac{\pi^2-4}{64}\frac{1}{\eta^2}+\frac{3\pi^2-16}{192}+ \nonumber\\
&& \left(\frac{-17}{180}+\frac{13\pi^2}{960}\right)\eta^2+
\left(\frac{-383}{3780}+\frac{683\pi^2}{60480}\right)\eta^4+...\Bigg]+... \; .
\label{eqd0} 
\end{eqnarray}
Equation (\ref{eqd0}) confirms that, by construction, we have
\begin{equation}
F^{GAP}= F^{Lind}, \;\;\rm{when}\;\;\Delta=0\ .
\label{eq9}
\end{equation}
Inspection of Eqs. (\ref{eqe0}) and (\ref{eqd0}) clearly shows that
\begin{eqnarray}
\lim_{\Delta\rightarrow0}\lim_{\eta\rightarrow0}F^{GAP} & = & \infty\ , 
\label{eq10} \\
\lim_{\eta\rightarrow0}\lim_{\Delta\rightarrow0}F^{GAP} & = &1\ , \label{eq10a} 
\end{eqnarray}
meaning that $F^{GAP}$ has an ``order of limits problem''. 
Such a situation is common in DFT. For example, we recall 
that several meta-GGA XC functionals (e.g. TPSS \cite{TPSS}, revTPSS 
\cite{perdew2009workhorse,perdew2011erratum}, BLOC \cite{bloc,blochole}, SA-TPSS \cite{constantin2016semilocal}, VT\{8,4\} 
\cite{del2012new}) suffer of such a 
order of limits problem. Nonetheless, they are accurate for many systems and properties, 
showing realistic system-averaged XC hole models \cite{blochole}. 

In the opposite limit, i.e. for $\eta\rightarrow \infty$, we have
\begin{eqnarray}
&& F^{GAP}\rightarrow 
3\eta^2-\frac{3}{5}+ \nonumber\\
&& (-\frac{24}{175}+\frac{3}{16}\Delta^2)\frac{1}{\eta^2}+
\mathcal{O}(\frac{1}{\eta^4}) \,. 
\label{eq8}
\end{eqnarray}
Therefore, in this limit, $F^{GAP}$ always behaves as $F^{Lind}$ for $\Delta=0$.

\begin{figure}[t]
\includegraphics[width=\columnwidth]{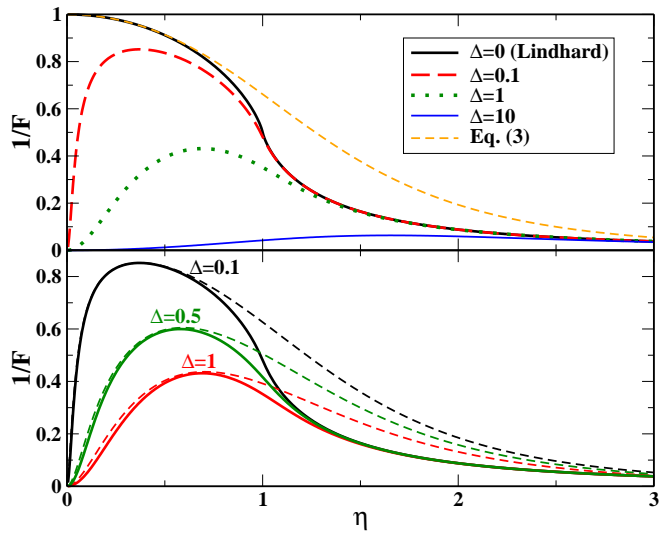}
\caption{Upper panel:$1/F^{GAP}$ versus $\eta$ for various values of $\Delta$. Also shown is the 
small-$\eta$ expansion of Eq. (\ref{eq2}). 
Lower panel: Comparison between $F^{GAP}$ (solid-lines) and the
expansion (dashed-lines) of Eq. (\ref{eqd0}), for $\Delta=0.1$, 0.5, and 1, respectively.
}
\label{f1}
\end{figure}

In the upper panel of Fig. \ref{f1}, we show $1/F^{GAP}$ for several values of $\Delta$. 
The plots are all smooth. 
At large $\eta$, $F^{GAP}$ recovers the Lindhard function [see Eq. (\ref{eq8})], while at small $\eta$ 
it is driven by the term $\propto \eta^{-2}$. 
The plot of the linear response of $T_s^{Lind4}[n]$ (Eq. (\ref{eq2})) is also given for comparison.
%
%
In the lower panel of Fig. \ref{f1} we report the accuracy of Eq. (\ref{eqd0}), considering only the terms
explicitly indicated in the equation, for $\Delta=0.1, 0.5,$ and  $1$.
Even for the case $\Delta=1$, this expansion is still very accurate for $\eta\leq 1$.

\subsection{Kinetic energy gradient expansions from  the linear-response of the jellium-with-gap model}
\label{subll1}
Next we proceed to build the linear-response jellium-with-gap KE gradient expansion, that should 
recover Eq. (\ref{eq4}) when $\Delta=0$. To this purpose, we consider the GAP4 expansion,
with the general form of the KE fourth-order gradient expansion
\begin{eqnarray}
&& T_s^{GAP4}[n]=\int d\R \tau^{TF}(\frac{a_1}{s^2}+\frac{a_2}{s}+a_3+a_4 s+ a_5 q+ \nonumber\\
&&+a_6 s^2+a_7 s q+a_8 s^3+a_{9}s^4+a_{10}q^2+a_{11} s^2 q) \; .
\label{gap4exp}
\end{eqnarray}
Performing the linear response of such a functional
\begin{equation}
F(\eta)=\frac{k_F}{\pi^2}\mathcal{F} \left( \frac{\delta^2 T_s[n]}{\delta n(\R)\delta n(\R')}|_{n_0}\right),
\label{eq14}
\end{equation}
where $\mathcal{F}$ represents the Fourier transform, we can find the coefficients $a_{i}$, by 
comparing term-by-term with Eq. (\ref{eqd0}). Nevertheless, the straightforward calculation of 
Eq. (\ref{gap4exp}) requires a tedious and long algebra \cite{nazarov2011optics,karimi2014three}.
Instead, a more elegant and simpler way to obtain the linear response of a given semilocal functional has 
been proposed in Ref. \onlinecite{tao2008nonempirical}: consider a small pertubation in density at 
$\R=\mathbf{0}$, of the form $n=n_0+n_ke^{i\mathbf{k}\R}$, such that $\nabla n=n_k i 
\mathbf{k}e^{i\mathbf{k}\R}$, and $\nabla^2 n=-n_k k^2e^{i\mathbf{k}\R}$, with $n_k\ll n_0$. Thus, at 
$\R=\mathbf{0}$, these expressions are simply $n=n_0+n_k$, $\nabla n=n_k i \mathbf{k}$, and $\nabla^2 n=-n_k k^2$. 
Inserting them in the functional expression, the linear response is obtained as twice the 
second-order coefficient of the series expansion with respect to $n_k/n_0$. After some algebra, 
the KE gradient expansion which gives the linear response of Eq. (\ref{eqd0}) is found to be
\begin{eqnarray}
&& T_s^{GAP4}[n]=\int d\R \tau^{TF}[\Delta^2 \frac{27}{91} \frac{\pi^2-4}{64}\frac{1}{s^2}+
\Delta\frac{5\pi}{72}\frac{1}{s}+
1+ \nonumber \\
&& \Delta^2(\frac{\pi^2}{64}-\frac{1}{12})+ 
 \Delta\frac{5\pi}{36} s+
(\frac{5}{27}+\Delta^2(\frac{-17}{324}+\frac{13\pi^2}{1728}))s^2+ \nonumber \\
&& \Delta\frac{-7\pi}{216} sq+ 
 (\frac{8}{81}+\Delta^2(\frac{-383}{6804}+\frac{683\pi^2}{108864}))q^2].
\label{gap4}
\end{eqnarray}  
The terms $\propto s^{-2}$ and $\propto s^{-1}$  account for the terms $\propto \eta^{-2}$ and 
$\propto \eta^{-1}$ of Eq. (\ref{eqd0}). These terms contribute only for a non-zero gap,
i.e. in semiconductors and insulators, but not in 
metals. At $\Delta=0$, $T_s^{GAP4}[n]$ correctly recovers $T_s^{Lind4}[n]$. 

To test $T_s^{GAP4}[n]$, we perform calculations for noble atoms, up to $Z=290$ electrons, 
using LDA orbitals and densities, in the Engel code \cite{engel1993accurate,engel2003orbital}.  
We consider $\Delta=2E_g/k_F^2(\R)$ with $E_g$ being the KS band gap of the atoms. Because the gradient
expansion is well defined only at small gradients and small-$\Delta$, we perform all the integrations 
over the volume $V$ defined by the conditions $-1\leq q \leq 1$ and $\Delta\leq 1$,
in a similar manner as in Ref. \onlinecite{constantin2016semiclassical}.
%
\begin{table*}
\begin{center}
\caption{\label{ta1} Comparison of several linear-response KE gradient expansions.
All integrations are performed over the volume $V$, defined by $-1\leq q \leq 1$ and $\Delta\leq 1$.
We show the exact KE ($T_s^{exact}$) and the errors $E_s^{approx}=T_s^{approx}-T_s^{exact}$ (in Hartree).
The GAP4 and LGAP functionals are defined in Eq. (\ref{gap4}) and Eq. (\ref{lgap}), respectively.
The best result of each line is shown in bold style. We use LDA orbitals and densities.}
\begin{ruledtabular}
\begin{tabular}{lrrrrr}
atom & $T_s^{exact}$ & $E_s^{GE2}$ & $E_s^{Lind4}$ & $E_s^{GAP4}$ & $E_s^{LGAP-GE}$ \\
\\
\hline
Ne           & 125.8 & {\bf -0.2} & 1.2 & 4.0 & 1.9 \\
Ar           & 512.2 & {\bf 4.3} & 8.1 & 12.0 & 11.4 \\
Kr           & 2742.3 & -21.1 & -6.7 & {\bf 1.0} & 10.3 \\
Xe           & 7214.4 & -50.7 & -19.6 & {\bf -8.9} & 23.2 \\
Rn           & 21829.6 & -146.5 & -72.4 & -56.8 & {\bf 48.9} \\
Uuo          & 46259.6 & -298.4 & -162.3 & -139.6 & {\bf 81.1} \\
168 $e^{-}$  & 106907.1 & -636.4 & -369.8 & -336.9 &  {\bf 158.9} \\
218 $e^{-}$  & 198077.5 & -1065.9 & -622.5 & -579.7  & {\bf 318.9} \\
290 $e^{-}$  & 389072.0 & -1888.9 & -1114.7 & -1056.4 & {\bf 630.1} \\
\end{tabular}
\end{ruledtabular}
\end{center}
\end{table*}
%
The results are reported in Table \ref{ta1}. For small atoms (Ne and Ar), the GE2 is 
more accurate than $T_s^{Lind4}[n]$ and $T_s^{GAP4}[n]$. However, we recall that in the case of a 
small number of electrons, the semiclassical and statistical concepts beyond the gradient 
expansions do not hold. In fact, for larger atoms (Kr to 
the noble atom with 290 electrons), both $T_s^{GAP4}$ and $T_s^{Lind4}$ outperform 
GE2. In particular, $T_s^{GAP4}$ shows the best performance, 
improving over $T_s^{Lind4}[n]$ and proving that, due to the inclusion of the gap, 
$F^{GAP4}$ contains important physics beyond $F^{Lind4}$.  

\subsection{Local band-gap}

In order to use Eq.~(\ref{gap4}) in semilocal DFT, we need to replace the true band gap $E_g$, with a density 
dependent 
local band gap. There are several models for the local band gap \cite{fabiano2014generalized,krieger1999electron},
constructed from the exponentially decaying density behavior \cite{krieger1999electron} or from conditions 
of the correlation energy \cite{fabiano2014generalized}. In the slowly-varying density limit, they behave as 
$E_g\sim |\nabla n|^m$, with $m\ge 2$. However, none of them can be considered accurate in this density regime. 

On the other hand, under a uniform density scaling $n_\lambda(\R)=\lambda^3 n(\lambda \R)$, 
the local band gap should behave as  $E_g\sim \lambda^2$. 
This condition is fulfilled by the general formula
\begin{equation}
E_g(\R)=a|\nabla n(\R)|^m/n(\R)^{2(2m-1)/3},\;\;\; m\ge 0, \;\;\; a\ge 0.
\label{eq16}
\end{equation}
Because other exact conditions of the local gap in the slowly varying density limit are not known, 
we use Eq. (\ref{eq16}) in the expression of $T_s^{GAP4}$, considering the case with $m=2$.
We fix the parameter $a$ requiring that the gradient expansion should recover the first two terms of the
kinetic energy asymptotic expansion for the large, neutral atom 
\cite{thomas1927calculation,fermi1927metodo,scottLEDPMJS52,schwingerPRA80,schwingerPRA81,englertPRA84,englertPRA85,elliottPRL08,   
lee2009condition,fabiano2013relevance}
\begin{equation}
T_s=c_0 Z^{7/3}+c_1 Z^2+c_2 Z^{5/3}+...,
\label{eq17}
\end{equation}
where $Z$ is the number of electrons. The first term in Eq. (\ref{eq17}) is the Thomas-Fermi one 
\cite{thomas1927calculation,fermi1927metodo}, the second 
is the Scott correction due to the atomic inner core \cite{scottLEDPMJS52}, 
and the last term accounts for quantum
oscillations \cite{schwingerPRA80,schwingerPRA81,englertPRA84,englertPRA85}. 
The exact coefficients are shown in the first line of Table 
\ref{ta2}. As in Ref. \onlinecite{lee2009condition}, we assume that any gradient expansion 
that is exact for the uniform electron gas, should 
have the exact $c_0$ coefficient. The calculation of $c_1$ and $c_2$ has been done using 
the method proposed in Ref. \onlinecite{lee2009condition}. 
We recall that the semiclassical atom theory has been often used in 
the development of exchange functionals 
\cite{constantin2016semiclassical,constantin2011semiclassical,sunPRL15,sunPNAS15,b88,cancio2012laplacian}
and occasionally also for kinetic energy functionals \cite{laricchia2011generalized}. 
Finally, we mention that 
these gradient expansions are models for the total KE, and not for the KE density, 
where the use of the reduced Laplacian $q$ (which is not 
present in linear response of the jellium model) is essential 
\cite{cancioJCP16,cancio2016visualizing,kinairy14,smiga2015subsystem}. 
%
\begin{table}
\begin{center}
\caption{\label{ta2} The coefficients $c_0$, $c_1$, and $c_2$ of the large-$Z$ expansion of 
the kinetic energy (see Eq. (\ref{eq17})).}  
\begin{ruledtabular}
\begin{tabular}{lrrr}
 & $c_0$ & $c_1$ & $c_2$  \\
\hline
Exact & 0.768745 & -0.500 & 0.270 \\
GE2 & 0.768745 & -0.536 & 0.336 \\
LGAP-GE & 0.768745 & -0.500 & 0.283 \\
\end{tabular}
\end{ruledtabular}
\end{center}
\end{table}
%


Using the procedure described above, we find $a=0.0075$, and we obtain the following gradient expansion (denoted as LGAP-GE)
\begin{eqnarray}
& T_s^{LGAP-GE}= \int d\R \tau^{TF}[1+a \frac{5\pi}{72}s+(\frac{5}{27}+a^2 \frac{27}{91}
\frac{\pi^2-4}{64})s^2+ \nonumber \\
& a \frac{5\pi}{36}s^3+\mathcal{O}(|\nabla n|^4)] \nonumber \\
& = \int d\R \tau^{TF}[1+0.0131 s+ 0.18528 s^2+0.0262 s^3]. 
\label{lgap}
\end{eqnarray}
Note that in Eq. (\ref{lgap}) only terms to up $s^3$ are considered (terms in Eq. (\ref{gap4}) proportional to $q$ or 
$q^2$ are neglected, as these terms will correspond to $s^4$).
 
As shown in Table \ref{ta2}, LGAP-GE gives a very accurate large-$Z$ expansion, having the $c_2$ coefficient 
close to exact. 
The results for noble atoms are reported in Table \ref{ta1}. LGAP-GE is reasonably accurate
for all atoms and, as expected due to the inclusion of the semiclassical atom theory, 
the accuracy increases with the number of electrons.

One additional
observation is that
LGAP-GE contains odd powers of the reduced gradient, in contrast with $F^{Lind4}$.
Nevertheless, Ou-Yang and Mel Levy have already shown that
using non-uniform coordinates scaling requirements
\cite{ou1990nonuniform}, the GE4 terms in the KE gradient expansion can be replaced by an $s$-only 
dependent term \cite{ou1991approximate}, whose
coefficient must be positive (and was fitted to the Xe atom). 
The resulting simple KE functional, that behaves better
than GE4 for the non-uniform density scaling, has the following enhancement factor
($F_s=\tau^{approx}/\tau^{TF}$):
\begin{equation}
F_s^{OL1}=1+\frac{5}{27}s^2+c s,
\label{eq19}
\end{equation}
with $c=0.01459$ being slightly bigger than its LGAP-GE counterpart.
Anyway, we need to acknowledge that, since
the kinetic potential of a GGA functional (with the enhancement factor $F_s$) has the general form
\begin{equation}
\frac{\delta T_s}{\delta n}=\frac{\partial \tau^{TF}}{\partial n}F_s(s)+\tau^{TF}\frac{\partial F_s}{\partial s}\frac{\partial 
s}{\partial n}-\nabla\cdot[\frac{1}{s}\frac{\partial F_s}{\partial s}\cdot\frac{\nabla n}{n^{8/3}}],
\label{eq20}
\end{equation}
a necessary condition for it to be well defined is $|\frac{1}{s}\frac{\partial F_s}{\partial s}|< \infty$.
This is not satisfied by 
the LGAP-GE (and OL1). Thus, the term $\propto s$ gives a diverging
kinetic potential ($\delta T_s/\delta n\rightarrow \infty$) at $s=0$. 
This is due to the high non-locality of Eq.
(\ref{gap4}), which was not fully suppressed by the local gap model of Eq. (\ref{eq16}) with $m=2$. 
Note that this divergence is a direct consequence of the jellium-with-gap theory.
Nevertheless, for molecular systems $s=0$ only at the middle of bonds, and it has been
found that this divergence is not important in real calculations of weakly-bounded molecular systems
\cite{gotz2009performance}. In fact, the same problem is shared by other well-known KE functionals
\cite{ou1991approximate,thakkar1992comparison,borgo2013density}. 

\section{Kinetic energy functional constructed from the LGAP gradient expansion}
\subsection{The LGAP GGA}
To show the importance of the LGAP-GE, we construct a simple GGA functional (named LGAP-GGA or simply LGAP) 
that recovers the LGAP-GE in the slowly-varying density regime. 
We consider the RPBE exchange enhancement factor form 
\cite{hammer1999improved}, $F_x^{RPBE}=1+\kappa(1-e^{-\mu s^2/\kappa})$, and we fix $\kappa=0.8$ from the Lieb-Oxford bound 
\cite{lieb1981improved}, using the approximate link between the kinetic and exchange energies 
(i.e. \textit{the conjointness conjecture} \cite{lee1991conjoint,march1991non,constantin2011semiclassical}). 
Note that, to our knowledge, the RPBE functional form has not been 
yet used in the development of kinetic functionals.
The LGAP kinetic enhancement factor is therefore defined as
\begin{equation}
F^{LGAP}_s=1 + \kappa \Big( 1 -  e^{-\mu_1 s-\mu_2 s^2-\mu_3 s^3} \Big),
\label{eq23}
\end{equation}
where $\mu_1=b_1/\kappa$, $\mu_2=b_2/\kappa+\mu_1^2/2$,
and $\mu_3=b_3/\kappa+\mu_1 \mu_2-\mu_1^3/6$,
such that it recovers the LGAP-GE in the slowly-varying density limit.
Here $b_1=0.0131$, $b_2=0.18528$, and $b_3=0.0262$ [see Eq. (\ref{lgap})].

\begin{figure}[t]
\includegraphics[width=\columnwidth]{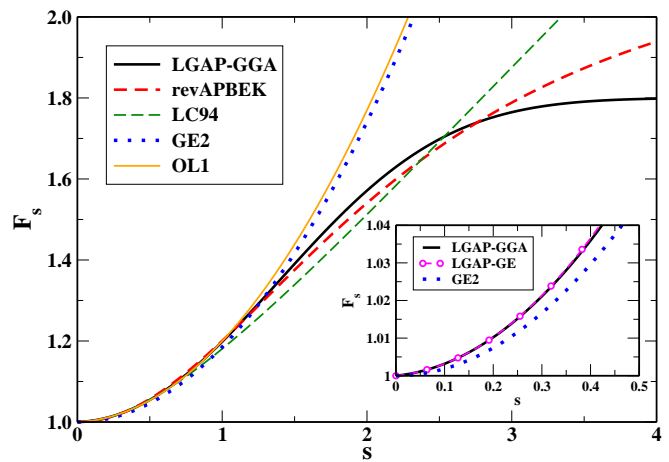}
\caption{Comparison of kinetic enhancement factors }
\label{fs}
\end{figure}
%
%
\begin{table*}
\begin{center}
\caption{\label{ta3} Mean absolute relative errors (MARE) of the non-self-consistent benchmark tests,
and mean absolute errors (MAE in mHa) of FDE self-consistent tests, given by several KE functionals. 
The best result of each group is highlighted in bold style.
}
\begin{ruledtabular}
\begin{tabular}{lrrrrr}
  & GE2 & revAPBEk & OL1 & LC94 &  LGAP \\ 
\hline
 \multicolumn{6}{c}{Total KE (non-self-consistent calculations)}\\
Atoms and ions           & 1.1 & 1.2 & 1.1& \bf{0.8} & 1.1 \\
Jellium clusters         & 1.0 & \bf{0.8} & 1.0 & 0.9 & \bf{0.8} \\
Jellium slabs            & 0.6 & 0.5 & 0.5& 0.5 & \bf{0.4} \\
Molecules                & 0.9 & 0.4 & 0.7 & 0.5 & \bf{0.2} \\
\multicolumn{6}{c}{KE differences (non-self-consistent calculations)}\\
Jellium cluster DKE      & 27.2 & 23.1 & 28.9 & \bf{21.3} & 22.6 \\
Jellium surfaces         & 3.3 & 3.6 & 3.4& 3.8 & \bf{3.1} \\
Jellium slabs dKE        & 5.0 & 3.5 & 4.7 & 4.1 & \bf{3.4} \\
Molecules AKE            & 184 & 155 & 185 & \bf{154} & 159 \\
\hline
\multicolumn{6}{c}{FDE results for molecular systems \footnotemark[1] (self-consistent calculations)}\\
Weak-interactions (WI)   & 2.46 & \textbf{0.13} & 2.49& 0.36 & 0.21 \\
Dipole interactions (DI) & 6.48 & 0.48 & 6.59& 0.67 & \textbf{0.45} \\
Hydrogen bonds (HB)      & 10.68 & \textbf{1.27} & 10.90& 1.34 &  1.69 \\
Dihydrogen bonds (DHB)   & 4.39 & 3.08 & 4.50& 2.92 & \textbf{2.58} \\
Charge transfer (CT)     & 5.05 & 2.61 & 6.94& 2.79 & \textbf{1.95} \\
MAE FDE                  & 5.66 & 1.72 &  6.31 & 1.82 & \textbf{1.50} \\
\end{tabular}
\end{ruledtabular}
\footnotetext[1]{Embedding energy errors 
$\Delta E = E^{\text{FDE}} -E^{\text{KS}} $ (mHa) for different KE functionals and complexes. 
In the last line, the mean absolute error (MAE) is reported.} 
\end{center}
\end{table*}
%
\subsection{The kinetic energy benchmark}
In order to assess the LGAP KE functional, we consider several known tests.

For {\bf total KE}:

\begin{itemize}
\item {The benchmark set of atoms and
ions \cite{perdew2007laplacian,laricchia2013laplacian,laricchia2011generalized}.
All calculations employed analytic Hartree-Fock orbitals and
densities \cite{CR74}};
\item {The Na jellium clusters ($r_s=3.93$) set with magic
electron numbers 2, 8, 18, 20, 34, 40, 58, 92, and 106, used in Refs.
\onlinecite{perdew2007laplacian,laricchia2013laplacian,laricchia2011generalized}. 
We use exact exchange orbitals and densities};
\item {The set of two interacting jellium slabs at different
distances \cite{laricchia2013laplacian}. 
Each jellium slab has $r_s=3$ and a thickness of $2\lambda_F$. Here $\lambda_F=2\pi/k_F$ is 
the Fermi wavelength. The calculations were performed using the orbitals and 
densities resulting from numerical Kohn-Sham calculations within the
local density approximation \cite{kohn1965self} for the XC
functional};
\item {The set of molecules 
(H$_2$, HF, H$_2$O, CH$_4$, NH$_3$, CO, F$_2$, HCN, N$_2$, CN, NO, and O$_2$) used in Refs.
\onlinecite{iyengar2001challenge,perdew2007laplacian,laricchia2013laplacian}. 
The noninteracting kinetic energies of test molecules
were calculated using the PROAIMV code \cite{KBT82}.
The required Kohn-Sham orbitals were obtained by Kohn-Sham calculations
performed with the uncontracted 6-311+G(3df,2p) basis set, the
Becke 1988 exchange functional \cite{b88}, and Perdew-Wang
correlation functional \cite{PW91}}.
\end{itemize}

For {\bf KE differences}: 

\begin{itemize}
\item The disintegration kinetic energy (DKE) of a jellium cluster \cite{constantin2009kinetic,laricchia2013laplacian};
\item The jellium surfaces test with bulk parameter $r_s$=2, 4, and 6 into the liquid drop model (LDM), as in Refs. 
\onlinecite{perdew2007laplacian,laricchia2013laplacian,laricchia2011generalized};
\item The dissociation KE (dKE) of a jellium slab into two pieces (as in Ref. \onlinecite{laricchia2013laplacian});
\item The atomization KE (AKE) of molecules \cite{iyengar2001challenge,perdew2007laplacian,laricchia2013laplacian}.
\end{itemize}


For {\bf non-additive KE}:
\vspace{0.2cm}

We employ the LGAP functional in subsystem DFT 
calculations, using the
TURBOMOLE \cite{turbomole} program, together with FDE script \cite{FDElar}.
The FDE calculations have been performed with a supermolecular def2-TZVPPD \cite{def2tzvpp,furchepol} basis set and the Perdew-Burke-Ernzerhof\cite{pbe} XC functional.
Five weakly interacting groups of molecular complexes are considered as a benchmark
\cite{laricchia2011generalized,laricchia2013laplacian,Laricchia2011114,savio2,savio3, smiga2015subsystem}:
\begin{itemize}
 \item[{\bf WI}:] weak interaction (He-Ne, He-Ar, Ne$_2$, Ne-Ar, CH$_4$-Ne,
C$_6$H$_6$-Ne, (CH$_4$)$_2$);
 \item[{\bf DI}:] dipole-dipole interaction  ((H$_2$S)$_2$, (HCl)$_2$, H$_2$S-HCl, CH$_3$Cl-HCl,CH$_3$SH-HCN, CH$_3$SH-HCl);
 \item[{\bf HB}:] hydrogen bond ((NH$_3$)$_2$, (HF)$_2$, (H$_2$O)$_2$, HF-HCN,
(HCONH$_2$)$_2$, (HCOOH)$_2$);
 \item[{\bf DHB}:] double hydrogen bond (AlH-HCl, AlH-HF, LiH-HCl, LiH-HF,
MgH$_2$-HCl, MgH$_2$-HF, BeH$_2$-HCl, BeH$_2$-HF);
 \item[{\bf CT}:] charge transfer (NF$_3$-HCN,C$_2$H$_4$-F$_2$,NF$_3$-HCN,
C$_2$H$_4$-Cl$_2$, NH$_3$-F$_2$, NH$_3$-ClF, NF$_3$-HF, C$_2$H$_2$-ClF,
HCN-ClF, NH$_3$-Cl$_2$, H$_2$O-ClF, NH$_3$-ClF).
\end{itemize}

\subsection{Results}

We compare our results with revAPBEk \cite{constantin2011semiclassical} and LC94 \cite{lc94} GGAs, 
which are considered state-of-the-art KE functionals for FDE \cite{laricchia2011generalized}, 
as well as with GE2 \cite{Kirz57,brack1976extended} and OL1 \cite{ou1991approximate}.
The KE enhancement factors of the considered functionals are reported in Fig. \ref{fs}.
In the inset of Fig. \ref{fs}, we show that LGAP and LGAP-GE are identical 
(by construction) at relatively small values of
the reduced gradient ($0\leq s \leq 0.5$), both differing significantly from the GE2 behavior. 
Consequently, LGAP shows a bigger enhancement factor than both LC94 and revAPBEk (i.e. 
$F_s^{LGAP}\ge\sim F_s^{revAPBEk}\ge\sim F_s^{LC94}$) when $s\leq 2.5$. Such a
feature has been proved to be essential for jellium surfaces \cite{airy3}. 
On the other hand, the LGAP enhancement
factor recovers its maximum value $F_s\rightarrow 1+\kappa$ at $s\approx 3$, faster than revAPBEk.

In Table \ref{ta3} we report the numerical results of all the tests. 
For total KE tests, LGAP gives the best overall performance, among the considered functionals, 
being the best for jellium clusters, jellium slabs and molecules. 
For KE differences LGAP is the most accurate for jellium surfaces and dissociation KE
of jellium slabs.
We also mention that LGAP performs reasonably well for all the other tests, 
being in line with revAPBEk. 

Finally, LGAP outperforms the other functionals for the FDE theory, being especially 
accurate for dipole-dipole, dihydrogen bond and charge transfer interactions. 
These latter results show that, in agreement with the finding of 
Ref. \onlinecite{gotz2009performance}, the divergence at $s=0$ of the 
LGAP kinetic potential is not important for calculations of weakly-bounded molecular systems. 
Moreover, results indicate that the LGAP-GE gradient expansion
can be successfully used in the kinetic energy functional construction, which perform relatively well in FDE theory.

\section{Conclusions}
In conclusion, we have investigated the linear response of the jellium-with-gap model, 
in the context of semilocal kinetic functionals. We have shown that the Levine and Louie \cite{levine1982new} 
analytical generalization of the Lindhard function ($F^{GAP}$) contains important physics beyond jellium model, 
and in particular we mention the following properties:

$(i)$ $F^{GAP}$ recovers the Lindhard function when the band gap is zero (i.e. $E_g=0$);

$(ii)$ $F^{GAP}$ has the correct behavior (see Eq. (4)) at small wave vectors, expressing the material-dependent 
constant $b$ in terms of the band gap; 

$(iii)$ In the regime of small band gap energy (i.e. $E_g \leq E_F$, with $E_F$ being the Fermi energy),
$F^{GAP}$ gives the GAP4 gradient expansion of the kinetic energy (see Eq. (17)), which is band-gap-dependent, and 
performs remarkably well in the atomic regions where the density varies slowly, improving 
over $T_s^{Lind4}$ of Eq. (7) (see Table I).  

These features show that $F^{GAP}$ should be further 
investigated and exploited in the field of non-local kinetic functionals 
\cite{wang2002orbital,huang2010nonlocal,wang1998orbital,shin2014enhanced,ho2008analytic,
alonso1978nonlocal,garcia1996nonlocal,chacon1985nonlocal,garcia1998nonlocal,garcia2008approach,
karasiev2014progress,karasiev2014finite,karasiev2015chapter,cangi2010leading,ribeiro2016leading}, and we will like to 
address this important issue in further work.

Finally, by considering a local band gap model, and a simple enhancement factor form, we 
have constructed the 
non-empirical 
LGAP GGA kinetic energy functional, derived from the linear response of the jellium-with-gap model (a.i. the GAP4 
gradient expansion). This functional 
showed the best performance in the context of FDE theory. Thus, it can be further used in real applications. 
\newline

\section*{Acknowledgements}
This work was partially supported by the National Science Center under 
Grant No.
DEC-2013/11/B/ST4/00771 and DEC-2016/21/D/ST4/00903.

\bibliography{lr16sub}

\end{document}